\documentclass [a4paper,12pt]{article}
\usepackage{graphicx}
\usepackage[american]{babel}

\begin{document}

\title{Synchronization of forced quasi-periodic coupled
oscillators}
\author{Kuznetsov A.P., Sataev I.R., Turukina L.V.}

\maketitle

\begin{center}
\textit{ Kotel'nikov's Institute of Radio-Engineering and Electronics of RAS, Saratov Branch, \\ 
Saratov, Zelyenaya, 38, 410019, Russian Federation} \\
e-mail:lvtur@rambler.ru
\end{center}

\begin{abstract}
The problem of synchronization of coupled self-oscillators by
external force is studied. The charts of Lyapunov's exponents in
the "frequency - amplitude" parameter plane are obtained within
the framework of the phase approximation. We identified the
characteristic configurations of the domains of complete
synchronization, two- and three-frequency quasi-periodic
oscillations and different variants of partial synchronization of
oscillators by an external force. The differences between regimes
of mode locking and beats of partial oscillators are revealed and
discussed. To visualize and analyze the domains of resonant
two-frequency tori we construct the charts, in which regions of
dynamics with different winding numbers are represented by colors.
\end{abstract}



\section{Introduction}

In nonlinear science, there is a fundamental problem concerning
creation and evolution of multi-frequency quasi-periodic
oscillations. Such phenomena are common in nature and technology,
including electronics, laser physics, mechanics, as well as
neurodynamics, biochemistry, climatology
\cite{l1}-\cite{l5}.

In due time, a scenario of onset of hydrodynamic turbulence was
advanced by Landau and Hopf \cite{l6}, \cite{l7} based on the idea
of sequential appearance of increasing number of oscillatory
components with incommensurate frequencies due to successive
bifurcation in the course of increase of a control parameter (the
Reynolds number). Latter, it was criticized and reconsidered by
Ruelle and Takens \cite{l8}. They asserted that after a
birth of a few oscillatory modes the motion will be associated
with a strange attractor characterized by sensitivity of orbits to
initial conditions and continuous spectrum. Although the Ruelle -
Takens motivation was based on rigorously proven mathematical
theorem, its application to concrete systems with complex dynamics
in physics, technology, and other disciplines remains
questionable. Particularly, many researchers outline typical
occurrence of multi-frequency oscillations in various examples of
nonlinear systems \cite{l9}-\cite{l17}. One must
acknowledge that at the moment we do not have a complete and
satisfactory general picture of emergence and disappearance of
multi-frequency quasiperiodic motions, the role of synchronization
and chaos in this picture, illustrative visualization of
respective phenomena etc.

A convenient class of models for discussion of many aspects of the
problem is represented by ensembles composed of oscillators
(autonomous or non-autonomous). It is remarkable that novel and
unexpected features of dynamical behavior of such models were
found recently even in the simplest case of two coupled
self-oscillatory elements, say, when the degree of excess over the
thresholds of Andronov-Hopf bifurcations are essentially different
(see, e.g. \cite{l18}-\cite{l20}). The problem of the
dynamical behavior of three interacting oscillators is
considerably more complex and multifaceted. It can be studied in
the framework of different approaches: in terms of the original
differential equations of the system, on the level of
slow-amplitude description, and with the help of phase equations
and model maps for the phases (the maps on a torus). For example,
in Ref. \cite{l21} the model is discussed composed of
coupled maps of rotations. Within its framework a large number of
possible types of bifurcations were revealed, as well as the
possibility of occurrence of "toroidal" chaos was outlined. Note
that the problem of the dynamics of phases on the torus is complex
and deep; in particular, such thoughtful and non-trivial approach
like the renormalization group method was developed in this
context \cite{l22}, \cite{l23}. Theoretical and experimental
studies of three-frequency quasiperiodic motions in coupled
electronic oscillators were carried out e.g. in Refs.
\cite{l12}-\cite{l14}. One result is that emergence of
three-frequency oscillations is associated often with the
bifurcations of merging and disappearance of pairs of stable and
unstable two-frequency tori. In Ref. \cite{l15} the
authors investigated a model of two van der Pol oscillators with
reactive coupling excited by an external periodic force. For weak
coupling the dominant three-frequency quasiperiodicity was
observed, and for large couplings chaos becomes possible and
typical. Dynamics of a ring of three coupled phase oscillators was
discussed e.g. in Refs. \cite{l24}-\cite{l26}. A possibility
of three-frequency quasiperiodic motions was noticed in the model
of three-coupled Lorenz systems \cite{l16}, \cite{l17}, and its
destruction accompanied with transitions to chaos was considered.

In recent years, the problem of forced synchronization of two
oscillators by an external force has been studied in some details.
In a series of papers \cite{l27}-\cite{l31}
mechanisms for synchronization of resonance cycles on a torus in
the autonomous systems were established and discussed. In
\cite{l30}, \cite{l31} phase equations describing the excitation
of two coupled dissipative oscillators by periodic driving force
were obtained and analyzed. A picture of domains of distinct
dynamical behavior on the parameter plane of the frequency and
amplitude was revealed; it contains e.g. regimes of complete
locking of the oscillators by the external force, and regimes of
two-frequency and three-frequency quasiperiodic motions. This
result, to some extent, generalizes the concept of Arnold's
tongues on the situations of multi-frequency motions.

In the present paper, we study a model of periodically driven
system composed of two coupled van der Pol oscillators, and
corresponding phase equations. In comparison with
\cite{l27}-\cite{l31}, one novel respect concerns
consideration of regimes associated not only with the mode
locking, but as well with beats of the partial oscillators
subjected the action of the periodic driving force. Bifurcation
analysis of the complete synchronization state is carried out on
the base of relatively simple model describing phase dynamics of
the coupled oscillators. We perform analysis of the spectrum of
Lyapunov's exponents in the parameter plane of frequency mismatch
and amplitude of the external force, and summarize it by means of
visualization with the so-called "charts of tori". On such a
chart, the resonant two-frequency tori with different winding
numbers, which occur on the surface of a three-frequency torus,
are shown in colors. With this technique, we detect arrangement of
the synchronization domains in the parameter plane ignificantly
different for two situations: when the isolated two oscillators
are mode locked, and when they demonstrate the beats. Finally, we
compare and discuss typical Lyapunov charts in the parameter
planes plotted for the approximate phase equations and for the
original differential equations.

\section{Equations for the periodically driven phase oscillators}

Let us consider a system of equations describing dynamics of two dissipative
coupled van der Pol oscillators under external harmonic force acting on the
first oscillator:
\begin{equation}
\label{eq1}
\begin{array}{c}
 \ddot {x} - (\lambda - x^{2})\dot {x} + (1 - \displaystyle\frac{\Delta}{2})x + \mu
(\dot {x} - \dot {y}) = B \sin( \omega t), \\
 \ddot {y} - (\lambda - y^{2})\dot {y} + (1 + \displaystyle\frac{\Delta}{2})y + \mu
(\dot {y} - \dot {x}) = 0. \\
\end{array}
\end{equation}
Here $\lambda$ is the parameter responsible for the Andronov --
Hopf bifurcation in the uncoupled oscillators; $\Delta$ is the
detuning parameter, proportional to a difference of the oscillator
frequencies; $\mu$ is the coupling parameter; $B$ is the amplitude
of the external force and $\omega=1+\Omega$ is its frequency.
Central (mean) oscillator frequency is assumed to be normalized to
unity. Hence, $\Omega$ is the dimensionless detuning of the
frequency of external force from the central frequency.

We use the method of slow amplitudes \cite{l1} and present
the solution for dynamical variables $x$ and $y$ in the form:
\begin{equation}
\label{eq2} x = a e^{i \omega t} + a^{\ast} e^{ - i \omega t}, \quad
y = c e^{i \omega t} + c^{\ast} e^{ - i \omega t}
\end{equation}
where $a(t)$ and $c(t)$ are complex amplitudes of the oscillators
and $\omega$ is the external forcing frequency. Further we use the
additional conditions standard for the method of slow amplitudes:
\begin{equation}
\label{eq3} \dot{a} e^{i \omega t} + \dot{a}^{\ast} e^{ - i \omega
t} = 0, \quad \dot{c} e^{i \omega t} + \dot{c}^{\ast} e^{ - i \omega
t} = 0.
\end{equation}
Substitution of (\ref{eq2}) and (\ref{eq3}) into Eqs.(\ref{eq1}),
with multiplication by $e^{ - i \omega t}$ and subsequent time
averaging leads to a set of truncated equations:
\begin{equation}
\label{eq4}
\begin{array}{c}
 2\dot{a} = \lambda a - {\left| {a} \right|}^{2}a - 2i(\Omega +
\displaystyle\frac{\Delta}{4})a - \mu (a - c) - \displaystyle\frac{B}{4}, \\
 2\dot {b} = \lambda c - {\left| {c} \right|}^{2}c - 2i(\Omega -
\displaystyle\frac{\Delta}{4})c - \mu (c - a).
\end{array}
\end{equation}
In the last equation we took into account that $\omega=1$.
Parameter $\lambda$ may be eliminated by renormalization.

Further let us set $a=Re^{i \psi_{1}}$ and $c=re^{i \psi_{2}}$,
where $R, r$ are real amplitudes and $\psi_{1}, \psi_{2}$ are real
phases of the oscillators determined relatively to the external
signal. Following \cite{l1}, we assume in the last
equation that oscillators are moving in the neighborhood of the
limit cycle $R=r=1$. Finally we obtain the equations:
\begin{equation}
\label{eq5}
\begin{array}{c}
  \dot{\psi_{1}}= -\displaystyle\frac{\Delta}{4} -\Omega +
  \displaystyle\frac{\mu}{2} \sin(\psi_{2}-\psi_{1}) +b \sin(\psi_{1}),\\
  \dot{\psi_{2}}= \displaystyle\frac{\Delta}{4} -\Omega
  + \displaystyle\frac{\mu}{2} \sin(\psi_{1}-\psi_{2}),
\end{array}
\end{equation}
where $b=\displaystyle\frac{B}{4}$. These are phase equations for
the periodically driven coupled oscillators \cite{l31}.

In the case of uncoupled oscillators ($\mu=0$) the system
(\ref{eq5}) splits into two independent equations: phase equation
for the periodically driven first oscillator \cite{l1} and phase
equation for the autonomous second oscillator. If the external
force is absent one can introduce relative phase of the
oscillators $\theta=\psi_{1}-\psi_{2}$. In this case we obtain the
classical Adler equation for two dissipative coupled oscillators
$\dot{\theta}=-\displaystyle\frac{\Delta}{2}-\mu \sin(\theta)$
\cite{l1}. This equation describes phase locking for the relative
phase of the oscillators at $|\Delta| < 2\mu$ and beats at
$|\Delta| > 2\mu$. Phase locking of the oscillators in the case
when external driving is absent occurs at there central frequency.

\section{Complete frequency locking of oscillators by the external
force}

Now let us discuss the parameter plane
($\Omega, b$) for the system (\ref{eq5}). First we shall find the
domain of complete synchronization that corresponds to the exact
phase locking of the both oscillators
($\dot{\psi_{1}}=\dot{\psi_{2}}=0$). Substituting the last
condition into Eqs.(\ref{eq5}) and excluding
$\sin(\psi_{1}-\psi_{2})$ in the first equation we obtain:
\begin{equation}
\label{eq6}
\begin{array}{c}
  2\Omega=b \sin(\psi_{1}), \\
  2\Omega-\displaystyle\frac{\Delta}{4}=\mu \sin(\psi_{1}-\psi_{2}).
\end{array}
\end{equation}
The system (\ref{eq6}) defines the boundaries of the
synchronization region:
\begin{equation}
\label{eq7} b=\pm 2\Omega,
\end{equation}
\begin{equation}
\label{eq71} \Omega=\displaystyle\frac{\Delta}{4}\pm\frac{\mu}{2}.
\end{equation}

\begin{figure}[!ht]
\centering
\includegraphics[scale=0.4]{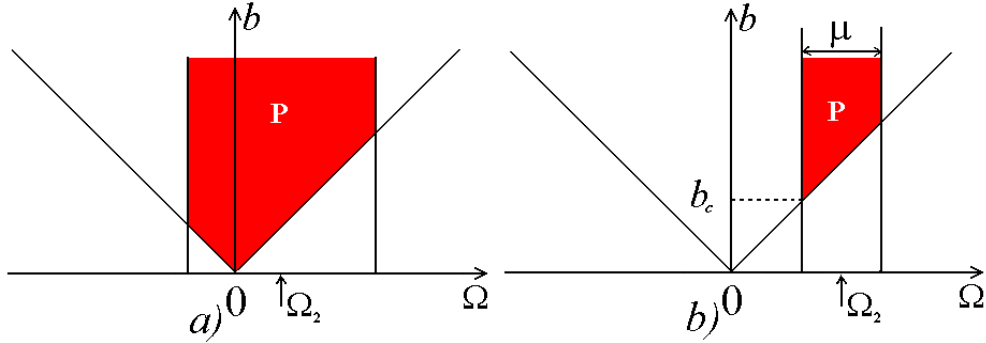}
\caption{\label{r1} Regions of phase locking (red
color) and quasi-periodic dynamics (white color) for the driven
system (\ref{eq5}) in the case of: a) phase locking of the
autonomous coupled oscillators; b) quasi-periodic dynamics of the
autonomous coupled oscillators. $\Omega_{2}$ is the intrinsic
frequency of the second oscillator; $b_{c}$ is the amplitude
threshold of synchronization of the quasi-periodic oscillations;
$\mu$ is the coupling parameter.}
\end{figure}

Condition (\ref{eq7}) defines the classical synchronization tongue
in the plane($\Omega, b$) (see Fig.\ref{r1}.a). The cusp is at the
point $\Omega=0$ which corresponds to the frequency of phase
locking of autonomous oscillators. However not all of the regimes
inside this synchronization domain are stable. Condition
(\ref{eq71}) supplements condition (\ref{eq7}) and determines the
width of a region of the stable synchronous regime. The center of this region coincides
with the intrinsic frequency of the second oscillator
$\Omega_{2}=\displaystyle\frac{\Delta}{4}$. And it's width is equal
to a coupling constant $\mu$ (Fig.\ref{r1}.a).

The boundaries of the region of complete synchronization by the
external force are given by the combination of the relations
(\ref{eq7}) and (\ref{eq71}). As a result, we obtain two types of
the synchronization regions presented in Fig.\ref{r1}.a and
Fig.\ref{r1}.b. From relation (\ref{eq7}) and (\ref{eq71}) it
follows that the first type of synchronization region is realized
at $|\Delta| < 2\mu$, and the second type at $|\Delta| > 2\mu$.
Hence, the first type of synchronization region corresponds to the
case when the oscillators in the absence of the external forcing
are phase locked, whereas the second type corresponds to the case
of quasi-periodic dynamics.

Comparison of Fig.\ref{r1}.a and Fig.\ref{r1}.b reveals some
typical features of the synchronization of oscillator in the
regime of beats. In this case an amplitude threshold for the
complete synchronization of the quasi-periodic oscillations
arises. From relations (\ref{eq7}) and (\ref{eq71}) we find it as:
\begin{equation}
\label{eq8} b_{c}=\displaystyle\frac{\Delta}{2}-2\mu.
\end{equation}
In accordance with relation (\ref{eq71}), the frequency of the
external force should be close to the intrinsic frequency of the
second oscillator $\Omega_{2}=\displaystyle\frac{\Delta}{4}$ to
obtain the complete synchronization of the quasi-periodic regime.

\section{Phase locking regime in autonomous oscillators}

\begin{figure}[!ht]
\centering
\includegraphics[scale=0.2]{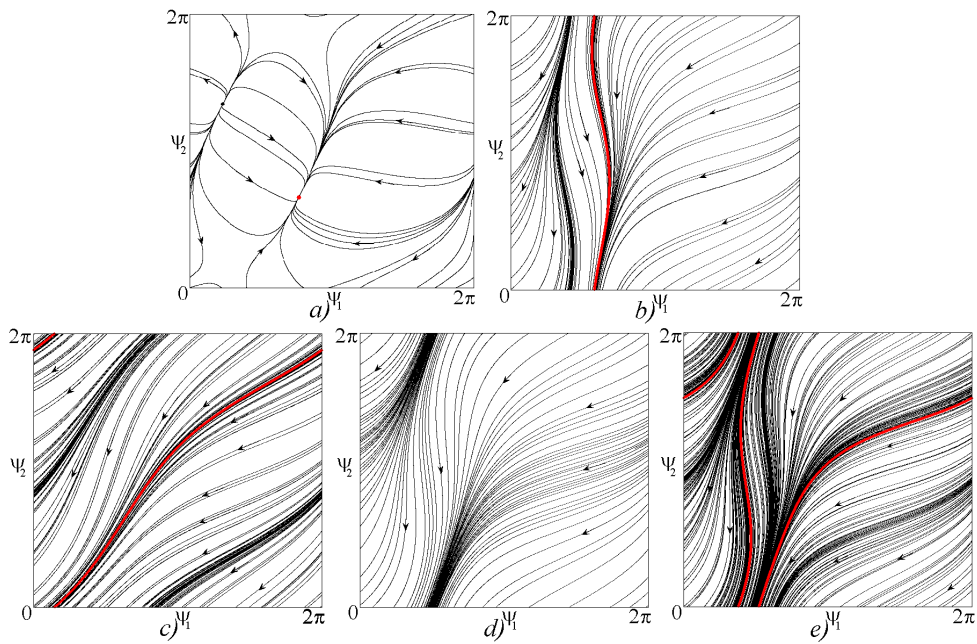}
\caption{\label{r2} Phase portraits of the system
(\ref{eq5}) at $\mu=0.3$ and $\Delta=0.2$: a) the complete
synchronization of both oscillators, $\Omega=0.1$, $b=0.3$; b) the
partial synchronization of phase of the first oscillator,
$\Omega=0.75$, $b=0.9$; c) the partial synchronization of the
relative phase of oscillators $\Omega=0.75$, $b=0.4$; d) the
three-frequency torus, $\Omega=0.75$, $b=0.75$; e) the resonant
two-frequency torus with winding number
$w=\displaystyle\frac{1}{2}$, $\Omega=0.75$, $b=0.79$. Attractors
are indicated by the red color.}
\end{figure}

System (\ref{eq5}) can demonstrate various types of dynamics
differing by the phase portrait on the ($\psi_{1}, \psi_{2}$)
plane. Typical phase portraits are shown in Fig.\ref{r2}. There
may be stable equilibrium points (Fig.\ref{r2}.a); various types
of attracting invariant curves (Fig.\ref{r2}.b,c,e); or
trajectories covering densely the whole phase plane
(Fig.\ref{r2}.d). These types of dynamical behavior correspond
respectively to the regimes of regular oscillations, two- and
three-frequency tori in the initial system.

We will use next method (method of the charts of Lyapunov's
exponents) for determining the type of the regime of system
(\ref{eq5}). We calculate all Lyapunov's exponents
\footnote{Checking the zero Lyapunov exponents was carried out
with an accuracy of $10^{-3}$. The accuracy of calculating the
main Lyapunov exponents was at least $10^{-4}$.} of the system
(\ref{eq5}) $\Lambda_{1}$, $\Lambda_{2}$ at each grid point in the
parameter plane ($\Omega, b$). Then we color the points of the
plane in accordance with values of the Lyapunov's exponents to
visualize the domains of the corresponding regimes:
\begin{enumerate}
  \item $P$ is the region of the stable equilibrium point (complete
phase locking), $\Lambda_{1}<0$, $\Lambda_{2}<0$;
  \item $T_{2}$ is the
region of two-frequency quasi-periodic regime, $\Lambda_{1}=0$,
$\Lambda_{2}<0$;
  \item $T_{3}$ is the
region of three-frequency quasi-periodic regime, $\Lambda_{1}=0$,
$\Lambda_{2}=0$ .
\end{enumerate}

Note that regime $T_{2}$ (Figs.\ref{r2}.b,c,e) corresponds to an attractor in the
form of two-frequency torus in the phase space of the original
system. Accordingly, three-frequency torus is a phase portrait for
regime $T_{3}$ (Fig.\ref{r2}.d).

The following number $w=\displaystyle\frac{p}{q}$ is useful to
classify the quasiperiodic regimes. Here $p$ and $q$ are numbers
of crossings of the phase trajectory in steady regime with the
sides of phase square ($0 < \psi_{1} < 2\pi$, $0 < \psi_{2} <
2\pi$). Number $w$ will be rational number (for which $p$ and $q$
are integer) for resonant two-frequency tori. In turn, for the
three-frequency tori $w$ is irrational \footnote{The value of $w$
can be interpreted as a kind of winding number. Indeed, closing
the region ($0 < \psi_{1} < 2\pi$, $0 < \psi_{2} < 2\pi$) for both
phase variables, we can obtain a "torus of phases", for which $w$ is
a winding number.}.

\begin{figure}[!ht]
\centering
\includegraphics[scale=0.45]{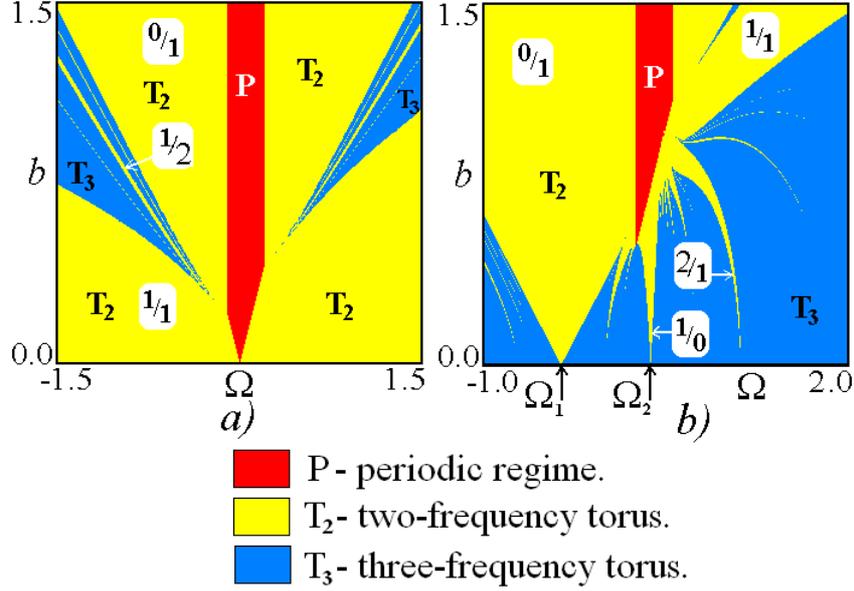}
\caption{\label{r3} Charts of Lyapunov's exponents
for the phase system (\ref{eq5}) for next cases: a) the phase
locking of the autonomous oscillators, $\mu=0.3$ and $\Delta=0.2$;
b) the quasi-periodical behavior of the autonomous oscillators,
$\mu=0.3$ and $\Delta=1.6$.}
\end{figure}

In Fig.\ref{r3}.a we present the charts of Lyapunov's exponents for
the system (\ref{eq5}). It is plotted in the case
when the autonomous oscillators are phase locked.  Domain of the periodical regimes $P$
presented in Fig.\ref{r3}.a corresponds to the analytical
investigation of the complete synchronization presented above.
System (\ref{eq5}) has three unstable and one stable equilibrium
points in this region. The stable point defines the regime of the
phase locking of coupled oscillators (Fig.\ref{r2}.a).

Two codimension-$2$ bifurcation point are disposed in ($\Omega,
b$) plane (Fig.\ref{r3}.a). Their coordinates are
\begin{equation}
\label{eq9} b=\displaystyle\frac{\Delta}{2} \pm \mu, \quad
\Omega=\displaystyle\frac{\Delta}{4} \pm \frac{\mu}{2}.
\end{equation}

The domains of two-frequency tori, three-frequency tori and
periodical regimes adjoin at these points \footnote{From the point
of theory of bifurcations two lines of saddle-node bifurcations
for stable points (\ref{eq7}), (\ref{eq71}) and lines of
saddle-node bifurcations for stable and unstable invariant curves
(tori) are merging at the mentioned points
[10,16-18,26]. Similar points for coupled
rotation maps in Ref.[10] were called "saddle-node fan", due to
the typical arrangement of resonant two-frequency tori of
different order, see fig.4.21 in Ref.[10].}. These points
determine the threshold of the amplitude of the signal at which
three-frequency regimes are possible. Regions of three-frequency
tori $T_{3}$ emerging from these points divide the domain of
two-frequency tori into two typical regions:
\begin{enumerate}
  \item Partial synchronization of the first oscillator by external force.
Winding number is $w=\displaystyle\frac{0}{1}$. In this case the
phase of the first oscillator $\psi_{1}$ is fluctuating around
some fixed value. While the phase of the second oscillator
$\psi_{2}$ demonstrates unlimited grow (Fig.\ref{r2}.b). This regime is observed
at large values of the parameter $b$.
  \item Partial synchronization of the relative phase of oscillators. Winding
number is $w=\displaystyle\frac{1}{1}$. In this case the phases of
both oscillators demonstrate unlimited grow. However, relative
phase of oscillators $\theta=\psi_{1}-\psi_{2}$ is fluctuating
around some fixed value (Fig.\ref{r2}.c). This regime is observed
at small values of the parameter $b$.
\end{enumerate}

The nature of these regimes is as
follows. Small external force cannot destroy the strong phase
locking of autonomous oscillators but it can synchronize both
oscillators as a whole. Large force can destroy synchronization
between the subsystems and then the first oscillator may be locked
by the external force. Note, that partial synchronization of
second oscillator is not possible. Second oscillator can only be
synchronized by external force simultaneously with the first
oscillator within the domain $P$.

In Fig.\ref{r3}.a one can see very narrow
regions inside the $T_{3}$ domain. Two-frequency tori are observed
in these narrow regions. These are the high order resonant 2D tori
situated on the surface of the corresponding three-frequency
torus. The phase portrait (Fig.\ref{r2}.e) corresponds to the
widest synchronization region. One can see that it is
two-frequency torus with winding number
$w=\displaystyle\frac{1}{2}$.

\section{Quasi-periodic regime in autonomous oscillators}

Let us increase detuning parameter $\Delta$ and pass to the
quasi-periodical regime in autonomous oscillators. Corresponding
chart of lyapunov's exponents is shown in Fig.\ref{r3}.b.
The domain $P$ of complete synchronization of both oscillators
by the external force corresponds to that shown in
Fig.\ref{r1}.b. Complete synchronization regime demonstrates
the threshold behavior with respect to the external force
amplitude. The value of this threshold is determined by the
formula (\ref{eq8}).

The region of three-frequency tori transforms essentially. The
amplitude threshold for 3D tori disappeared and they occupy the
lower part of the parameter plane ($\Omega, b$).

Region of the partial phase locking of the first oscillator
$w=\displaystyle\frac{0}{1}$ now looks like the traditional
synchronization tongue locating within the domain of the
three-frequency tori. The cusp of this synchronization region
is at the point $\Omega_{1}=-\displaystyle\frac{\Delta}{4}$ which
corresponds to the intrinsic frequency of the first oscillator.

Distinctive feature of this case is the
possibility of existence of two-frequency torus with winding
number $w=\displaystyle\frac{1}{0}$. It corresponds to partial
synchronization of second oscillator. Domain of this regime has a
form of the synchronization tongue whose top adjoins the region of
the exact phase locking $P$ while the cusp is at point
$\Omega_{2}=\displaystyle\frac{\Delta}{4}$ which corresponding
to the intrinsic frequency of the second oscillator. Thus, in the case
under consideration, the partial synchronization of the second
oscillator which is not directly excited by the external force is
possible. It constitutes important difference from the case when the
autonomous oscillators are phase locked. In the latter case the
coupling between oscillators is strong. And if external force has
frequency $\Omega \approx \Omega_{2}$ and its amplitude is small
then only the relative phase can be locked. But if the frequency
detuning $\Delta$ is large (the autonomous system of
coupled oscillators demonstrated the quasi-periodical regime) then the
second oscillator is not so strongly influenced by the first one.
In that case external force can lock the second oscillator, while
the phase of first oscillator is drifting.

Regime of partial locking of the relative phase of the oscillators
$w=\displaystyle\frac{1}{1}$ at $\Omega > 0$
is mostly separated from the frequency axes by the wide region of
three-frequency tori.

\section{Chart of two-frequency tori}

In Fig.\ref{r3} one can see a set of narrow synchronization
regions corresponding to different types of two-frequency tori. We
use next method for its detail investigation. Let us fix a point
in the parameter plane ($\Omega, b$). Then we solve numerically
the system (\ref{eq5}) for these parameter values, and find the
number of intersections of p and q of the phase trajectory with
the bottom and left sides of the phase square ($0 < \psi_{1} <
2\pi$,$0 < \psi_{2} < 2\pi$). (It is necessary to exclude
irrelevant crossings associated with the possible oscillatory
nature of the invariant curve.) Then we color the point in the
parameter plane in a certain color depending on the type of the
two-frequency torus which we can characterize by the value
$w=\displaystyle\frac{p}{q}$. Three-frequency tori are determined
as non-periodic regimes. To obtain a chart we perform a similar
procedure at each point of the grid in the parameters plane.

\begin{figure}[!ht]
\centering
\includegraphics[scale=0.45]{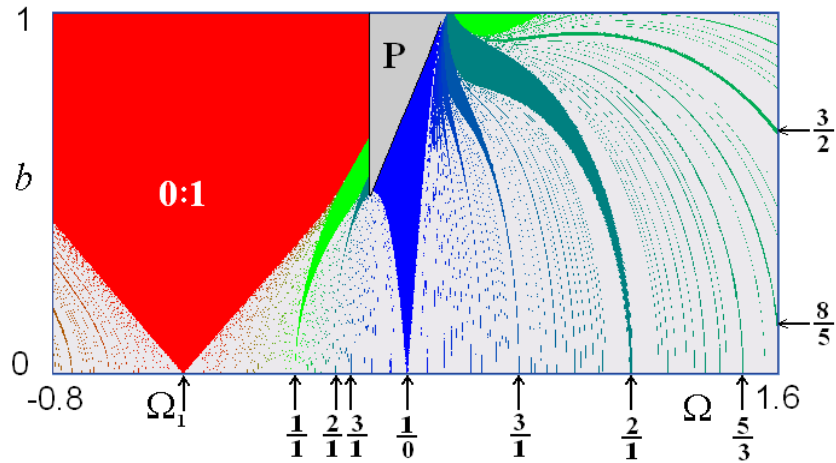}
\caption{\label{r4} Charts of the tori for the
case when the autonomous system demonstrates regimes of beats.
$\mu=0.3$ and $\Delta=1.6$. Numbers indicate the values of winding
number for the main two-frequency tori.}
\end{figure}

In Fig.\ref{r4} we present chart of tori plotted using a method
described above. This chart corresponds to the beats in the
autonomous oscillators. In this case the widest synchronization
region ($w=\displaystyle\frac{0}{1}$) corresponds to the partial
phase locking of the first oscillator. Its cusp is at the point at
$\Omega_{1}=-\displaystyle\frac{\Delta}{4}$ which corresponding to
the intrinsic frequency of the first oscillator. The domain of
partial phase locking of the second oscillator
($w=\displaystyle\frac{1}{0}$) looks like a continuation of the
region of complete phase locking $P$. Its cusp is at the point
$\Omega_{2}=\displaystyle\frac{\Delta}{4}$ which corresponds to
intrinsic frequency of the second oscillator. In turn, the
point $\Omega=\Omega_{2}$ is the center of whole system of the
regions of two-frequency tori of higher order.

\section{Dynamics of the original system}

\begin{figure}[!ht]
\centering
\includegraphics[scale=0.33]{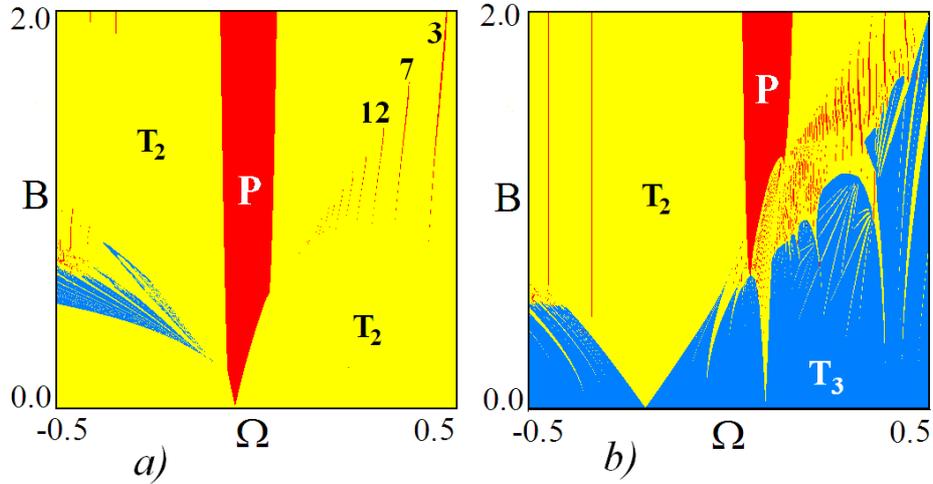}
\caption{\label{r5} Charts of Lyapunov's exponents
for the periodically driven system of coupled van der Pol
oscillators (\ref{eq1}): a) phase locking of autonomous
oscillators, $\mu=0.1$ and $\Delta=0.1$; b) beats of autonomous
oscillators, $\mu=0.1$ and $\Delta=0.6$.}
\end{figure}

Phase equations (\ref{eq5}) considered above are the approximation for the
original system (\ref{eq1}). They were obtained for the case of
$\lambda << 1$. Hence, it is important to clarify organization of
the "frequency - amplitude" parameter plane for system (\ref{eq1})
in the case when the value of the control parameter $\lambda$ is
not small. Charts of Lyapunov's exponents for system (\ref{eq1})
are shown in Fig.\ref{r5}. They are plotted for $\lambda=1$. Fig.\ref{r5}.a
corresponds to the case when the autonomous oscillators are phase
locked. And Fig.\ref{r5}.b corresponds to the case when the
autonomous system demonstrates beats.

Despite the large value of the control parameter
$\lambda=1$, the picture has much in common with the case of the
phase equations approximations. First, it concerns the
configuration of the region of complete synchronization $P$. It is
only slightly distorted as compared with Fig.\ref{r3}. In the case
when autonomous oscillators are phase locked one can see
codimension-two point in the region $\Omega < 0$ (Fig.\ref{r5}.a). Clearly visible
is a characteristic pattern of two- and three-frequency tori in
its vicinity. The corresponding point in the high-frequency domain
($\Omega > 0$) which can be seen in Fig.\ref{r3}.a disappears
together with the adjacent region of three-frequency tori.
Two-frequency tori and the system of very narrow regions of
periodical regimes of higher order are observed in its place.
These periodical regimes are absent in the phase equations.
Periods of some regimes are indicated in Fig.\ref{r5} by numbers.
These periods are defined numerically as the periods of cycles in
the stroboscopic section of the system (\ref{eq1}).

A chart plotted for the case of beats in autonomous oscillators
(Fig.\ref{r5}.b) also largely corresponds to the case of the phase
equations (Fig.\ref{r3}.b). First, we should note the presence of amplitude
threshold for the complete synchronization of quasi-periodic
oscillations. Also in the lower part of the chart one can see the
region of three-frequency tori, inside of which the regions of
different resonant two-frequency tori described in the previous
section are observed. However, the high-frequency codimension-two
point also disappears.

Another new effect in comparison with the case of phase equations
is an appearance at Fig.\ref{r5}.b of a set of additional "fans"
of the regions of two-frequency tori on the right side of the
chart. They are not associated with the region of complete
synchronization. These types of two-frequency tori are not
observed in the phase equations.

\begin{figure}[!ht]
\centering
\includegraphics[scale=0.2]{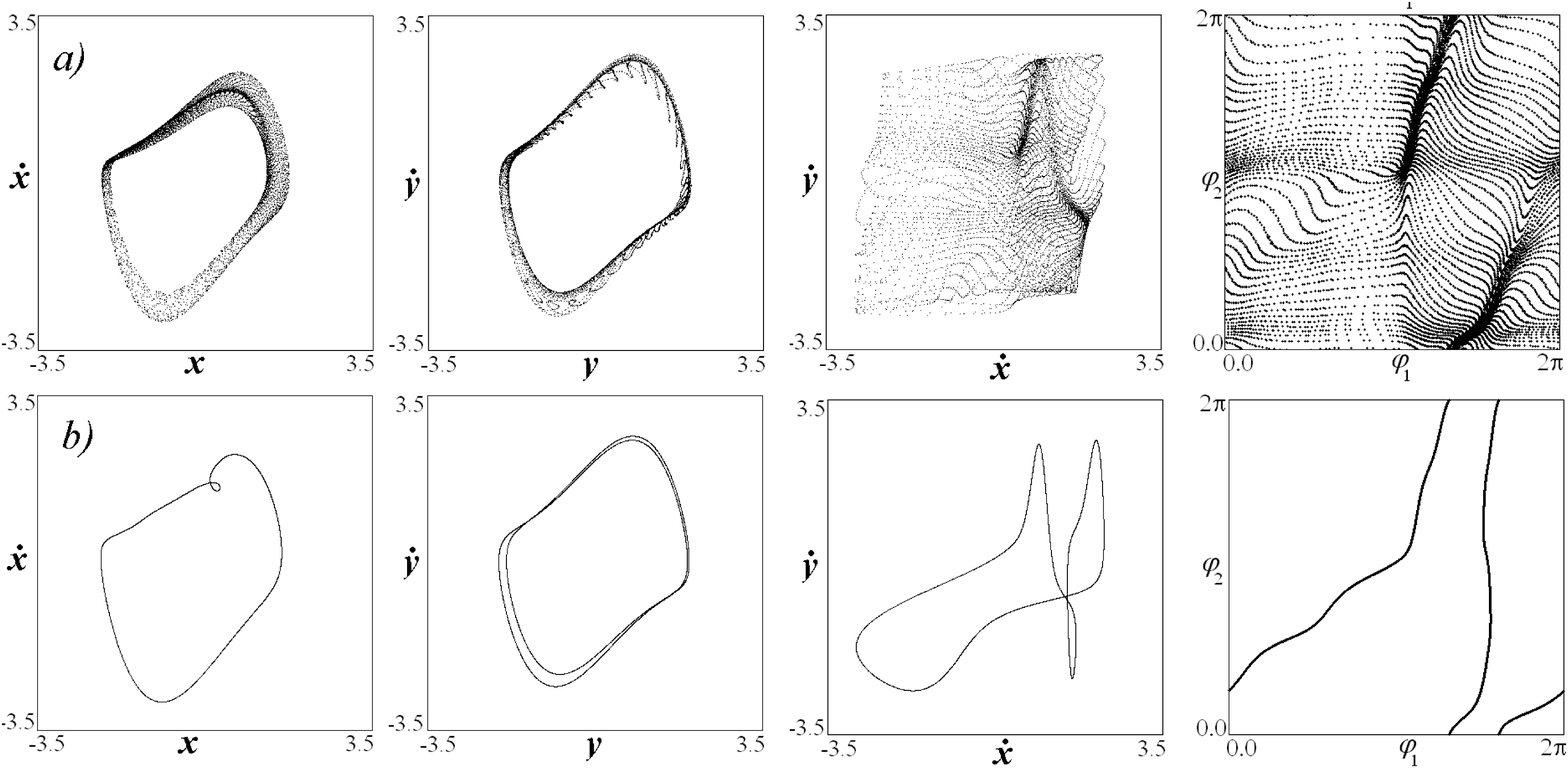}
\caption{\label{r6}Stroboscopic section of the forced system of
coupled van der Pol oscillators (\ref{eq1}) in the projection on
the plane of the variables of the first and second oscillators, in
the form of Lissajous figures, and portraits of the plane of the
oscillators phases in the stroboscopic section. Parameters are a)
$B=0.45$, b) $B=0.5$; $\lambda=1$, $\Delta=0.1$, $\mu=0.1$,
$\Omega=-0.25$.}
\end{figure}

Stroboscopic map is a convenient way to visualize the type of the
dynamical regime. It is a shift map along the trajectories of the
periodically driven system taken over the period of external
force. The result is an orbit - a set of points embedded in the
space of dynamical variables of the system (\ref{eq1}) ($x$,
$\dot{x}$, $y$, $\dot{y}$). For their visualization we use
projections on the plane of the dynamical variables of the first
and second oscillators ($x$, $\dot{x}$) and ($y$, $\dot{y}$).
These illustrations are given in the first two columns in
Fig.\ref{r6}. The third column shows the same projection on the
velocities plane of the two oscillators ($\dot{x}$, $\dot{y}$.) It
is a kind of Lissajou's figures. For the orbits from the first two
columns of Fig.\ref{r6} let us determine the phases of oscillators
$\varphi_{1}$ and $\varphi_{2}$, as the angles at which the points
of the orbit are visible from the origin. The trajectories on the
phase plane ($\varphi_{1}$, $\varphi_{2}$) obtained using this
method are presented in the fourth column in Fig.\ref{r6}.
Fig.\ref{r6}.a illustrates the three-frequency torus in the system
(\ref{eq1}). The trajectory in this case covers densely the plane
of the oscillators phases ($\varphi_{1}$, $\varphi_{2}$).
Fig.\ref{r6}.b illustrates a resonant two-frequency torus with
winding number $w=\displaystyle\frac{1}{2}$. This two-frequency
torus arises from a three-frequency torus with a small change in
the amplitude of the external force.

\section{Conclusion}

It is necessary to distinguish the cases of phase locking and quasi-periodic
dynamics with incommensurate frequencies in autonomous oscillators when we
investigate synchronization of two-frequency oscillations by the external
force. In the first case regimes of complete synchronization of two
oscillators, partial synchronization of relative phase of oscillators and
partial synchronization of first oscillator are prevailing. Appearing of
three-frequency tori is possible only above some external forcing amplitude
threshold. Simultaneous exact locking of phases of both oscillators by
harmonic external force is also possible in the case of quasi-periodic
dynamics of autonomous oscillators. However, now this effect demonstrates
amplitude threshold. There exists domain of regime of partial
synchronization of second oscillator as amplitude decrease: it has a form of
synchronization tongue in the parameter plane of frequency versus amplitude
which is surrounded by system of synchronization regions of resonant
two-frequency tori of high order. The amplitude threshold for
three-frequency tori disappears in that case. The method of charts of
dynamical regimes is an effective method for the analysis of such systems.
With this method we color the parameter plane in different colors in
accordance with winding number, which can be attributed to each resonant
two-frequency torus. This method may be used for the further analysis of
other similar systems. For increasing values of the control parameter of the
original system of coupled van der Pol equations the main features of the
organization of the parameters plane are preserved. In particular these are
the features characteristic for the transition from mode locking to the
beats in autonomous oscillators. However, at large amplitudes the
three-frequency tori disappear. In their place the domain of two-frequency
tori along with system of periodic regimes comes. The latter regimes have no
analogue in the phase equations.

{\it We thank S.P. Kuznetsov for help and discussions. This research
was supported by the grant RFBR No. 09-02-00426.}

\end{document}